\documentclass[12pt]{article}

\usepackage{pgf, tikz}
\usetikzlibrary{decorations.pathreplacing}
\usetikzlibrary{arrows}

\usepackage{tcolorbox}
\usepackage[utf8]{inputenc}
\usepackage[T1]{fontenc}
\usepackage[english]{babel}

\usepackage{xspace}
\usepackage{array}
\newcolumntype{L}[1]{>{\raggedright\let\newline\\\arraybackslash\hspace{0pt}}m{#1}}
\newcolumntype{C}[1]{>{\centering\let\newline\\\arraybackslash\hspace{0pt}}m{#1}}
\newcolumntype{R}[1]{>{\raggedleft\let\newline\\\arraybackslash\hspace{0pt}}m{#1}}

\usepackage{bm}
\usepackage[ruled,vlined,linesnumbered]{algorithm2e}
\usepackage{amsmath,amsfonts,amssymb,amsthm}
\usepackage{booktabs}
\usepackage{dsfont} 
\usepackage{multirow}
\usepackage{graphicx}
\usepackage{xcolor}
\usepackage{placeins}
\usepackage[top=90pt, bottom=90pt, left=75pt , right=75pt, headsep=12pt]{geometry}
\usepackage{pdf14}
\usepackage[pagebackref]{hyperref}
\newcommand{\mapolicebackref}[1]{\mbox{\textsl{\small #1}}}
\renewcommand*{\backref}[1]{}
\renewcommand*{\backrefalt}[4]{%
\ifcase #1 \mapolicebackref{Uncited in this paper}
    \or \mapolicebackref{#2}
    \else \mapolicebackref{#2}
\fi
}

\title{\titre}
\author{}

\newtheorem{definition}{Definition}[subsection]
\newtheorem{proposition}{Proposition}[subsection]
\newtheorem{theorem}{Theorem}[section]

\newtheorem{remark}[theorem]{Remark}

\newcommand{\ie}{\emph{i.e.~}}

\newcommand{\lar}{\stackrel{\mathdollar}{\leftarrow}}

\newcommand{\KeyGen}{\normalfont\textsf{KeyGen}}
\newcommand{\Encrypt}{\normalfont\textsf{Encrypt}}
\newcommand{\Decrypt}{\normalfont\textsf{Decrypt}}

\newcommand{\param}{\ensuremath{\mathsf{param}}}	
\newcommand{\Setup}{\ensuremath{\mathsf{Setup}}}

\newcommand{\x}{\bf x}

\newcommand{\sk}{\ensuremath{\mathsf{sk}}\xspace}

\newcommand{\pk}{\ensuremath{\mathsf{pk}}\xspace}

\newcommand{\word}[1]{\ensuremath{\boldsymbol{#1}}}

\newcommand{\cv}{\word{c}}

\newcommand{\ev}{\word{e}}

\newcommand{\rv}{\word{r}}

\newcommand{\xv}{\word{x}}
\newcommand{\yv}{\word{y}}
\newcommand{\zv}{\word{z}}

\usepackage{marvosym}
\usepackage{wasysym}
\usepackage{pdfpages}

\usepackage{arydshln}

\date{}

\usepackage{listings}
\definecolor{dkgreen}{rgb}{0,0.6,0}
\definecolor{gray}{rgb}{0.5,0.5,0.5}
\definecolor{mauve}{rgb}{0.58,0,0.82}

\title{HQC-RMRS, an instantiation of the HQC encryption framework with
a more efficient auxiliary error-correcting code}

 \author{Nicolas Aragon\thanks{XLIM, Universit\'e de Limoges} \and Philippe Gaborit\thanks{XLIM, Universit\'e de Limoges} \and Gilles Z\'emor \thanks{IMB, Universit\'e de Bordeaux}}
 \date{}

\begin{document}
\maketitle
\begin{abstract}
The HQC encryption framework is a general code-based encryption scheme for which
decryption returns a noisy version of the plaintext. Any instantiation of the
scheme will therefore use an error-correcting procedure relying on a fixed
auxiliary code. 
Unlike the McEliece encryption framework whose security is directly related to
how well one can hide the structure of an error-correcting code, 
the security reduction of the HQC encryption framework is independent of the
nature of the auxiliary decoding procedure which is publicly available.
What is expected from it is that the decoding algorithm is both efficient and has a decoding failure rate which can be easily modelized and analyzed. 
The original error-correction procedure proposed for the HQC framework was 
to use tensor products of BCH codes and repetition codes. 
In this paper we consider another code family for removing the error vector 
deriving from the general framework: 
the concatenation of Reed-Muller and Reed-Solomon codes. 
We denote this instantiation of the HQC framework by HQC-RMRS.
These codes yield better decoding results than the BCH and repetition codes: overall we gain roughly 17\% in the size of the key and the ciphertext, 
while keeping a simple modelization of the decoding error rate. The paper also
presents a simplified and more precise analysis of the distribution of the error
vector output by the HQC protocol. 
\end{abstract}

\section{Introduction}

The first code-based cryptosystem was proposed in 1978 by McEliece. 
The proposed framework can be instantiated with any family of error-correcting
codes having an efficient decoding algorithm. However, the security of the
cryptosystem is highly dependent on the choice of the family. Even though the
original instantiation, based on Goppa codes, remains secure, many others
(for example using Reed-Muller or Reed-Solomon codes) have been broken by
recovering the hidden structure of the code from the public key. 
The very nature of this framework makes it difficult to reduce the security of the scheme to a general decoding problem for random codes.

In \cite{ABDGZ18}, the authors propose a framework that can be instantiated in
different metrics to build cryptosystems whose security is reduced to decoding random quasi-cyclic codes. 
The instantiation of this framework in the Hamming metric is HQC (Hamming Quasi Cyclic), which was submitted to the second round of the NIST Post-Quantum Standardization workshop. Thanks to the quasi-cyclic structure, the scheme features compact key sizes (about 20,000 bits for a security of 128 bits) as well as fast keygen, encryption and decryption operations.

A public structured error-correcting code $\mathcal{C}$ is needed to remove
noise inherent to the decryption process. In \cite{ABDGZ18}, the authors
proposed tensor products of BCH and repetition codes, because encoding is fast
and they allow precise DFR analysis. The analysis consists of two steps: first,
the weight distribution of the error vector is studied,
and then one analyzes how well the chosen codes decode errors of a given weight.

Our contribution in this paper is twofold: we provide a better analysis of the
distribution of the weight of the error vector, which allows for a better DFR
analysis regardless of which public code is used to decode it, 
and we propose using a concatenation of Reed-Muller and Reed-Solomon codes to decode
the error: this code family allows one to reach a low DFR (for example $< 2^{-128}$) with shorter codes, hence leading to shorter public keys and ciphertexts.

\section{Preliminaries}

In this section we introduce necessary notation and the description of the HQC scheme. For more details about the protocol and the security proof, we refer the reader to \cite{ABDGZ18}.
\smallskip

\textbf{Notation:} Throughout this document, $\mathbb{Z}$ denotes the ring of integers and $\mathbb{F}_2$ the binary field. Additionally, we denote by $\omega(\cdot)$ the Hamming weight of a vector \ie{} the number of its non-zero coordinates, and by $\mathcal{S}_{w}^n\left(\mathbb{F}_2\right)$ the set of words in $\mathbb{F}_2^n$ of weight $w$. Formally:
$$\mathcal{S}_{w}^n\left(\mathbb{F}_2\right) = \left\lbrace\mathbf{v}\in \mathbb{F}_2^{n}\textnormal{, such that }\omega(\mathbf{v})=w\right\rbrace.$$
$\mathcal{V}$ denotes the vector space $\mathbb{F}_2^n$ of dimension $n$ 
over $\mathbb{F}_2$ for
some positive $n \in \mathbb{Z}$. Elements of $\mathcal{V}$ can be
interchangeably considered as row vectors or polynomials in
$\mathcal{R}=\mathbb{F}_2[X]/(X^{n}-1)$. Vectors/Polynomials (resp. matrices)
will be represented by lower-case (resp. upper-case) bold letters. A prime
integer $n$ is said primitive if the polynomial $X^n-1/(X-1)$ is irreducible in
$\mathbb{F}_2[X]$.

For $\mathbf{u, v} \in \mathcal{V}$, we define their product similarly as in $\mathcal{R}$, \emph{i.e.} $\mathbf{u}\mathbf{v} = \mathbf{w} \in \mathcal{V}$ with
\begin{equation}
\label{eq:product}
w_k = \sum_{i+j\equiv k \mod~n} u_i v_j\textnormal{, for }k \in \{0,1, \ldots, n-1\}.
\end{equation}

Our new protocol takes great advantage of the cyclic structure of matrices.
Following~\cite{ABDGZ18}, \textbf{rot}$(\mathbf{h})$ for $\mathbf{h} \in \mathcal{V}$ denotes the circulant matrix whose $i^{\textnormal{th}}$ column is the vector corresponding to $\mathbf{h}X^i$. This is captured by the following definition.

\begin{definition}[Circulant Matrix]
Let $\mathbf{v} = \left(v_0, \ldots, v_{n-1}\right) \in \mathbb{F}_2^n$. The \emph{circulant matrix} induced by $\mathbf{v}$ is defined and denoted as follows:
\begin{equation}\label{eq:rot}
\textnormal{\textbf{rot}}(\mathbf{v}) = \begin{pmatrix}
v_0 & v_{n-1} & \ldots & v_1\\
v_1 & v_0 & \ldots & v_2\\
\vdots & \vdots & \ddots & \vdots\\
v_{n-1} & v_{n-2} & \ldots & v_0\\
\end{pmatrix} \in \mathbb{F}_2^{n \times n}
\end{equation}
\end{definition}

As a consequence, it is easy to see that the product of any two elements $\mathbf{u, v} \in \mathcal{R}$ can be expressed as a usual vector-matrix (or matrix-vector) product using the \textbf{rot}$(\cdot)$ operator as
\begin{equation}
\mathbf{u}\cdot\mathbf{v} = \mathbf{u}\times\textnormal{\textbf{rot}}(\mathbf{v})^\top = \left(\textnormal{\textbf{rot}}(\mathbf{u})\times\mathbf{v}^\top\right)^\top = \mathbf{v}\times\textnormal{\textbf{rot}}(\mathbf{u})^\top = \mathbf{v}\cdot\mathbf{u}.
\end{equation}

We now recall the HQC scheme in figure \ref{fig:scheme}. In \cite{ABDGZ18}, the
code $\mathcal{C}$ used for decoding is a tensor product of BCH and repetition
codes. But since this code is public, its structure has no incidence on
security, and one can choose any code family, influencing only
the Decryption Failure Rate and the parameter sizes.

\begin{figure}[!ht]
\centering
\fbox{
\begin{minipage}{.965\textwidth}
\begin{itemize}
\item \Setup$(1^\lambda)$: generates and outputs the global parameters \param{} = $(n, k, \delta, w, w_\mathbf{r}, w_\mathbf{e})$.
\item \KeyGen$(\param)$: samples $\mathbf{h} \lar \mathcal{R}$, the generator matrix $\mathbf{G}\in \mathbb{F}_2^{k\times n}$ of $\mathcal{C}$, $\sk = (\mathbf{x}, \mathbf{y}) \lar \mathcal{R}^2$ such that $\omega(\mathbf{x})= \omega(\mathbf{y}) = w$, sets $\pk = \left(\mathbf{h}, \mathbf{s} = \mathbf{x+h\cdot y}\right)$, and returns $(\pk, \sk)$.
\item \Encrypt$(\pk, \mathbf{m})$: generates $\mathbf{e} \lar \mathcal{R}$, $\mathbf{r} = (\mathbf{r}_1, \mathbf{r}_2) \lar \mathcal{R}^2$ such that $\omega(\mathbf{e})=w_\mathbf{e}$ and $\omega(\mathbf{r}_1)=\omega(\mathbf{r}_2)=w_\mathbf{r}$, sets $\mathbf{u} = \mathbf{r}_1+\mathbf{h}\cdot\mathbf{r}_2$ and $\mathbf{v} = \mathbf{m}\mathbf{G} + \mathbf{s\cdot r}_2 + \mathbf{e}$, returns $\mathbf{c} = \left(\mathbf{u}, \mathbf{v}\right)$.
\item \Decrypt$(\sk, \mathbf{c})$: returns $\mathcal{C}$.$\mathsf{Decode}(\mathbf{v}-\mathbf{u\cdot y})$.
\end{itemize}
\end{minipage}
}
\caption{\label{fig:scheme}Description of \textsf{HQC}.}
\end{figure}

\section{Analysis of the error vector distribution for Hamming distance}
\label{sec:error_distribution}

From the description of the HQC framework, decryption corresponds to decoding the received vector: ${\bf v-u.y=mG+\mathbf{e}^\prime}$ for the error vector ${\bf e^\prime = \mathbf{x\cdot r}_2 - \mathbf{r}_1\cdot \mathbf{y} + \mathbf{e}}$. In this section we provide a more precise analysis of the error distribution approximation compared to \cite{ABDGZ18}. 
We first compute exactly the probability distribution of each fixed coordinate
$e_k^\prime$ of the error vector
 \[
\mathbf{e}^\prime = \mathbf{x\cdot r}_2 - \mathbf{r}_1\cdot \mathbf{y} +
\mathbf{e}=(e_0',\ldots e_{n-1}').
\]
We obtain that every coordinate $e'_k$ is Bernoulli distributed
with parameter $p^*=P[e_k^\prime=1]$ given by Proposition~\ref{pr:error-distro}.

To compute decoding error probabilities, we will then need the probability distribution
of the weight of the error vector $\mathbf{e}^\prime$
restricted to given sets of coordinates that
correspond to codeword supports. We will make the simplifying assumption that
the coordinates $e_k'$ of $\mathbf{e}^\prime$ are independent variables, which
will let us work with the binomial distribution of parameter $p^*$ for the
weight distributions of $\mathbf{e}^\prime$. This working assumption is
justified by remarking that, in the high weight regime relevant to us, since the
component vectors $\mathbf{x},\mathbf{y},\mathbf{e}$ have fixed weights, the
probability that a given coordinate $e_k'$ takes the value $1$ conditioned on
abnormally many others equalling $1$ can realistically only be $\leq p^*$.
We support this modeling of the otherwise intractable weight distribution of
$\mathbf{e}^\prime$ by extensive simulations: these back up our assumption that
our computations
of decoding error probabilities and DFRs can only be upper bounds on their real
values.

\subsection{Analysis of the distribution of the product of two vectors}

\noindent The vectors $\mathbf{x},\mathbf{y},\mathbf{r}_1,\mathbf{r}_2,
\mathbf{e}$
have been taken uniformly random and independently chosen among vectors
of weight $w$, $w_\mathbf{r}$ and $w_\mathbf{e}$.
We first evaluate the distributions of the products $\mathbf{x\cdot
 r}_2$ and $\mathbf{r}_1\cdot \mathbf{y}$.

\begin{proposition}\label{prop:ptilde}
Let $\xv=(x_0,\ldots x_{n-1})$ be a random vector  chosen uniformly among all binary
vectors of weight $w$ and let 
$\rv=(r_0,\ldots ,r_{n-1}$) be a
random vector chosen uniformly among all 
vectors of weight $w_r$ and independently of $\xv$.
Then, denoting $\zv = \xv \cdot \rv$, we have that for every $k\in\{0,\ldots
n-1\}$, the $k$-th coordinate $z_k$ of $\zv$ is Bernoulli distributed with
parameter $\tilde{p}=P(z_k=1)$ equal to:

$$\tilde{p} = \frac{1}{{n \choose w}{n \choose w_{\mathbf{r}}}}
\sum\limits_{\substack{1 \leqslant \ell \leqslant \min(w, w_{\mathbf{r}})\\ \ell\
\mathrm{odd}}} C_\ell $$
where $C_\ell = {n \choose \ell}{n-\ell \choose w-\ell}{n-w \choose
w_{\mathbf{r}}-\ell}$.
\end{proposition}

\proof
The total number of ordered pairs $(\xv,\rv)$ is
$\binom{n}{w}\binom{n}{w_\mathbf{r}}$. Among those, we need to count how many
are such that $z_k=1$. We note that
\[
z_k=\sum_{\substack{i+j=k \bmod n\\ 0\leq i,j\leq n-1}}x_ir_j.
\]
We need therefore to count the number of couples $(\xv,\rv)$ such that 
we have $x_ir_{k-i}=1$ an odd number of times when $i$ ranges over $
\{0,\ldots ,n-1\}$ (and $k-i$ is understood modulo $n$).
Let us count the number $C_\ell$ of couples $(\xv,\rv)$ such that
$x_ir_{k-i}=1$ exactly $\ell$ times. For $\ell>\min(w,w_\mathbf{r})$ we clearly
have $C_{\ell}=0$. For $\ell\leq\min(w,w_\mathbf{r})$ we have $\binom{n}{\ell}$
choices for the set of coordinates $i$ such that $x_i=r_{k-i}=1$, then
$\binom{n-\ell}{w}$ remaining choices for the set of coordinates $i$ such that
$x_i=1$ and $r_{k-i}=0$, and finally $\binom{n-w}{w_\mathbf{r}-\ell}$ remaining
choices for the set of coordinates $i$ such that $x_i=0$ and $r_{k-i}=1$.
Hence $C_\ell=\binom{n}{\ell}\binom{n-\ell}{w-\ell}\binom{n-w}{w_{\mathbf{r}}-\ell}$.
The formula for $\tilde{p}$ follows.
\qed

\subsection{Analysis of $\ev'$}

Let $\mathbf{x, y}$ (resp. $\mathbf{r}_1, \mathbf{r}_2$) be independent random vectors
chosen uniformly among all binary
vectors of weight $w$ (resp. $w_\mathbf{r}$).

By independence of $(\mathbf{x},\mathbf{r}_2)$ with $(\mathbf{y},\mathbf{r}_1)$,
the $k$-th coordinates of $\mathbf{x\cdot r}_2$
and of $\mathbf{r}_1\cdot \mathbf{y}$ are independent, and they are Bernoulli
distributed with parameter $\tilde{p}$ by Proposition~\ref{prop:ptilde}.
Therefore their modulo $2$ sum
$\mathbf{t} = \mathbf{x\cdot r}_2 - \mathbf{r}_1\cdot \mathbf{y}$ is
Bernoulli distributed with
\begin{equation}\label{eq:sum-prod}
    \begin{cases}
     \mathrm{Pr}[t_k=1] =
      2\tilde{p}(1-\tilde{p}), \\
     \mathrm{Pr}[t_k=0] =
(1-\tilde{p})^2+\tilde{p}^2 .
    \end{cases}
\end{equation}

Finally, by adding modulo $2$ coordinatewise the two independent vectors $\mathbf{e}$ and $\mathbf{t}$,
we obtain the distribution of the coordinates of the error vector
$\mathbf{e}^\prime = \mathbf{x\cdot r}_2 - \mathbf{r}_1\cdot \mathbf{y} +
\mathbf{e}$ given by the following proposition:

\begin{proposition}\label{pr:error-distro}
Let 
$\xv,\yv$ be uniformly chosen among vectors of weight $w$, let $\rv_1,\rv_2$ be
uniformly chosen among vectors of weight $w_\mathbf{r}$, and let $\ev$ be
uniformly chosen among vectors of weight $w_\mathbf{e}$. We suppose furthermore
that the random vectors $\x_v,\yv,\rv_1,\rv_2,\ev$ are independent.
Let $\mathbf{e}^\prime = \mathbf{x\cdot r}_2 - \mathbf{r}_1\cdot \mathbf{y} +
\mathbf{e}=(e_0',\ldots ,e_{n-1}')$. Then, for every $k\in\{0,\ldots ,n-1\}$ we
have:
\begin{equation}
\label{eq:pr-total}
    \begin{cases}
\mathrm{Pr}[e^\prime_k=1] = 
2\tilde{p}(1-\tilde{p})(1-\frac{w_\mathbf{e}}{n}) +
\left((1-\tilde{p})^2+\tilde{p}^2\right)\frac{w_\mathbf{e}}{n}, \\
\mathrm{Pr}[e^\prime_k=0] = \left((1-\tilde{p})^2+\tilde{p}^2\right)(1-\frac{w_\mathbf{e}}{n}) +
2\tilde{p}(1-\tilde{p})\frac{w_\mathbf{e}}{n}.
    \end{cases}
\end{equation}
\end{proposition}

\proof
The vectors $\mathbf{x\cdot r}_2$, $\mathbf{r}_1\cdot \mathbf{y}$ and $\ev$ are
clearly independent. The $k$-th coordinate of $\ev$ is Bernoulli distributed
with parameter $w_\mathbf{e}/n$. The random Bernoulli variable $e_k'$ is
therefore the sum modulo $2$ of three independent Bernoulli variables of
parameters $\tilde{p}$ for the first two and of parameter $w_\mathbf{e}/n$ for
the third one. The formula therefore follows standardly.
\qed

\medskip

Proposition~\ref{pr:error-distro} gives us the probability that a
coordinate of the error vector $\mathbf{e}^\prime$ is $1$.
In our simulations, which occur in the regime $w=\alpha\sqrt{n}$ with constant
$\alpha$, we make the simplifying assumption that the coordinates
of $\mathbf{e}^\prime$ are independent, meaning that the weight of $\mathbf{e}^\prime$
follows a binomial distribution of parameter $p^\star$,
where $p^\star$ is defined as in Eq. (\ref{eq:pr-total}): $p^\star = 2\tilde{p}(1-\tilde{p})(1-\frac{w_\mathbf{e}}{n}) +
\left((1-\tilde{p})^2+\tilde{p}^2\right)\frac{w_\mathbf{e}}{n}$. This
approximation will give us, for $0\leq d \leq \min(2\times w\times w_\mathbf{r} +w_\mathbf{e}, n)$,
\begin{equation}\label{eq:weight}
\mathrm{Pr}[\omega(\mathbf{e}^\prime) = d] = \binom{n}{d}{(p^\star)}^d{(1-p^\star)}^{(n-d)}.
\end{equation}

\begin{figure}[h!]
\centering
\includegraphics{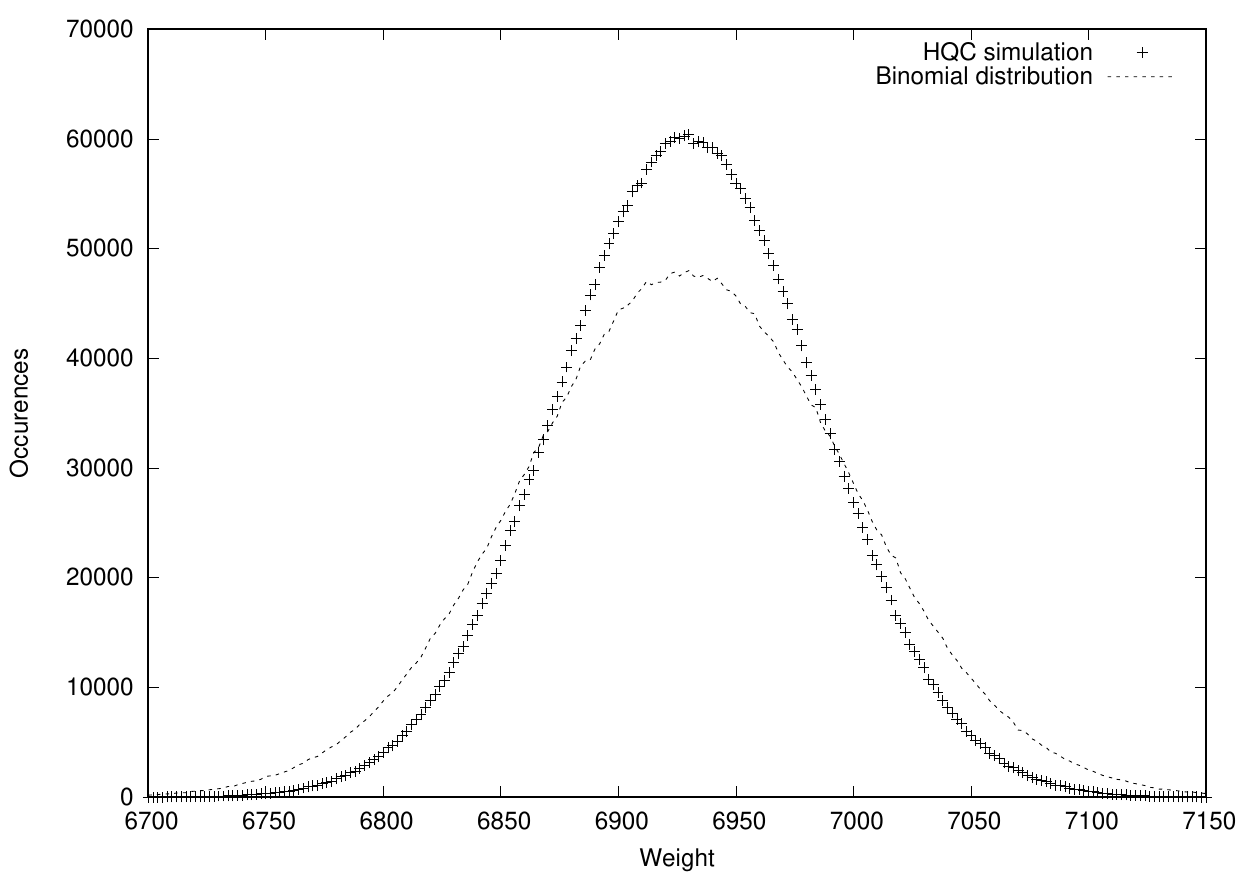}
\caption{Comparaison between error $\mathbf{e}^\prime$ generated using parameter set I and its binomial approximation.}
\label{fig:128_1_distribution}
\end{figure}

\begin{figure}[h!]
\centering
\includegraphics{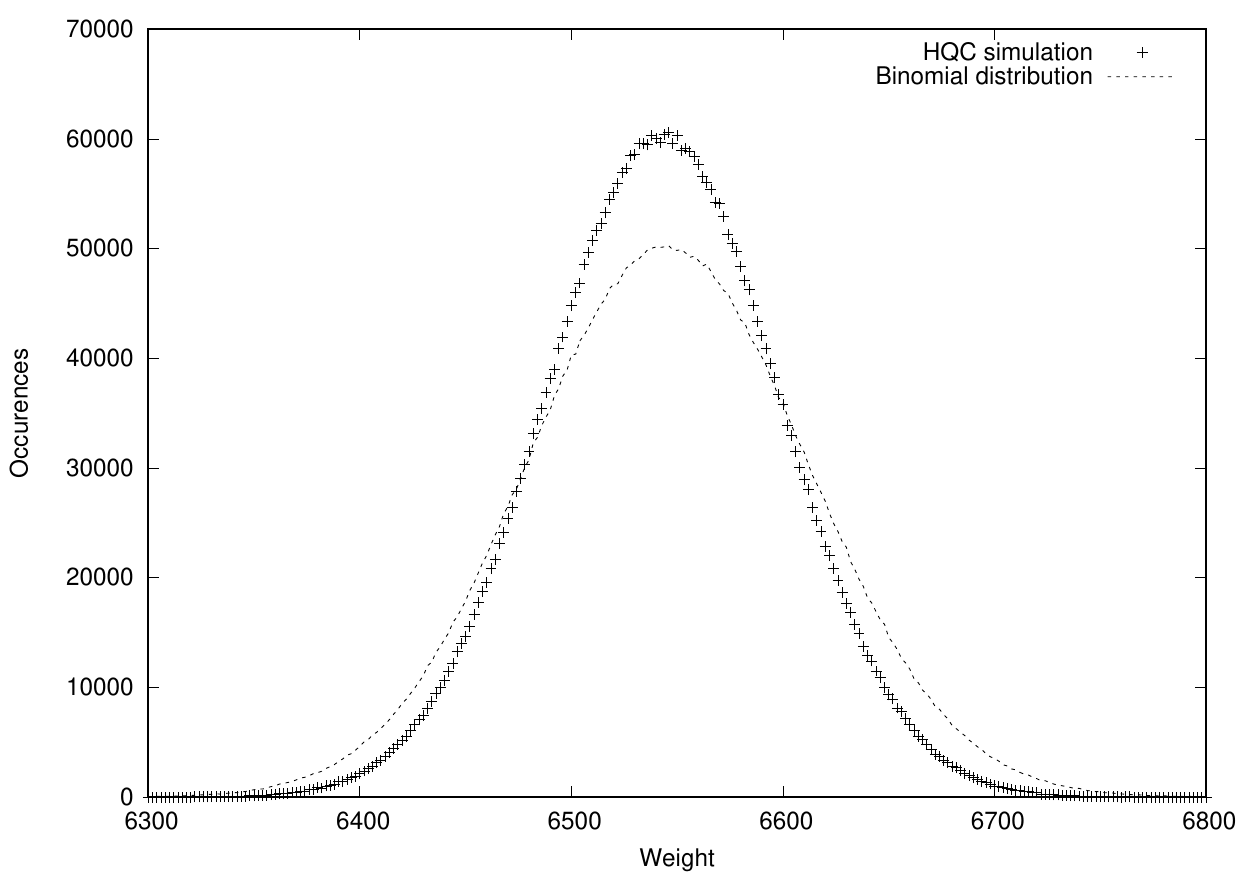}
\caption{Comparaison between error $\mathbf{e}^\prime$ generated using parameter set II and its binomial approximation.}
\label{fig:128_2_distribution}
\end{figure}

\subsection{Supporting elements for our modelization}

we give in Fig. \ref{fig:128_1_distribution} and \ref{fig:128_2_distribution}
simulations of the distribution of the weight of the
error vector together with the distribution of the associated binomial law of
parameters $p^\star$. These simulations show that error vectors
are more likely to have a weight close to the mean than
predicted by the binomial distribution,
and that on the contrary the error is less likely to be of large weight than if
it were
binomially distributed.
This is for instance illustrated on parameters sets I and II corresponding to
real parameters used for 128 bits security. For cryptographic purposes we are mainly interested by very small DFR and large weight occurences which are more likely to induce decoding errors. These tables show that the probability of obtaining a large weight is close but smaller for the error weight distribution of $e'$ rather than for the binomial approximation. This supports our modelization and the fact that computing the decoding failure probability with this binomial approximation permits to obtain an upper bound on the real DFR. This will be confirmed in the next sections by simulations with real weight parameters (but smaller lengths).

\bigskip

\noindent {\bf Examples of simulations.} We consider two examples of parameters in Table 1: Parameter sets I and II which correspond to cryptographic parameters and for which we simulate the error distribution versus the binomial approximation together with the probability of obtaining large error weights. In order to
follow \cite{ABDGZ18} we computed vectors of length $n$ (the blocksize of the double circulant code) and then, for $n_1n_2$ the length of the auxiliary error correcting code $C$, we truncated the last $l = n - n_1n_2$ bits before measuring the Hamming weight of the vectors.

\bigskip
\begin{table}[h]
\begin{center}
\begin{tabular}{|c|c|c|c|c|c|}
\hline
Parameter set & $w$ & $w_{\mathbf{e}} = w_{\mathbf{r}}$ & $n$ & $n_1n_2$ & $p^\star$\\
\hline
I & 67 & 77 & 23,869 & 23746 & 0.2918\\
\hline
II & 67 & 77 & 20,533 & 20480 & 0.3196\\
\hline
\end{tabular}
\end{center}
\caption{Parameters sets I and II used for simulations.}
\end{table}

\bigskip

\noindent \textbf{Simulations for Parameter set I.} Simulation results are shown figure \ref{fig:128_1_distribution}. We computed the weights such that 0.1\%, 0.01\% and 0.001\% of the vectors are of weight greater than this value, to study how often extreme weight values occur in Table 2.

\bigskip

\begin{table}[h]
\begin{center}
\begin{tabular}{|c|c|c|c|c|}
\hline
 & 0.1\% & 0.01\% & 0.001\% & 0.0001\%\\
\hline
 \hline
 Error vectors & 7101 & 7134 & 7163 & 7190\\
 \hline
 Binomial approximation & 7147 & 7191 & 7228 & 7267\\
 \hline
\end{tabular}
\end{center}
\caption{Simulated probabilities of large weights for Parameter Set I for the distributions of the error vector and the binomial approximation.}
\end{table}

\bigskip

\noindent \textbf{Simulations for Parameter set 2.} Simulation results are shown on figure \ref{fig:128_2_distribution}. We perform the same analysis as for the parameter set I about extreme weight values in Table 3.

Simulation results are shown on figure \ref{fig:128_2_distribution}. We perform the same analysis as for the parameter set I about extreme weight values in Table 2.

\bigskip

\begin{table}[h]
\begin{center}
\begin{tabular}{|c|c|c|c|c|}
\hline
 & 0.1\% & 0.01\% & 0.001\% & 0.0001\%\\
\hline
 \hline
 Error vectors & 6715 & 6749 & 6779 & 6808\\
 \hline
 Binomial approximation & 6753 & 6796 & 6834 & 6859\\
 \hline
\end{tabular}
\end{center}
\caption{Simulated probabilities of large weights for Parameter Set II for the distributions of the error vector and the binomial approximation.}
\end{table}

As we can see from these, extreme weight values seem to happen more often in the
case of the binomial approximation. Since these cases are the most likely to
lead to decoding failure, this approximation should lead to conservative decryption failure rate estimations.

\bigskip

\noindent {\bf Comparison with the previous analysis in \cite{ABDGZ18}:} the present analysis is
better than the previous one, in practice in the case of decoding with BCH and
repetition codes for security parameter 128 bits, the present analysis leads to
a DFR in $2^{-154}$ when the previous one lead to $2^{-128}$. In practice this
allows to reduce by 3\% the key size in the case of the BCH-repetition code decoder of \cite{ABDGZ18}.

\section{Proposition of new auxiliary error-correcting codes: Reed-Muller and Reed-Solomon concatenated codes}
\label{sec:concatenated}

In this section we study the impact of using a new family of auxiliary
error-coeecting codes: instead of the tensor product codes used in the HQC
cryptosystem framework we propose to consider the concatenation of Reed-Muller and Reed-Solomon codes. We denote this instanciation of the HQC framework by HQC-RMRS.

\subsection{Construction}

\begin{definition}{\bf [Concatenated codes]}

A concatenated code consists of an external code $[n_e, k_e, d_e]$ over $\mathbb{F}_q$ and an internal code $[n_i, k_i, d_i]$ over $\mathbb{F}_2$, with $q = 2^{k_i}$. We use a bijection between elements of $\mathbb{F}_q$ and the words of the internal code, this way we obtain a transformation:

$$ \mathbb{F}_q^{n_e} \rightarrow \mathbb{F}_2^N$$

where $N = n_e n_i$. The external code is thus transformed into a binary code of parameters $[N = n_e n_i, K = k_e k_i, D \geqslant d_e d_i]$.
\end{definition}

For the external code, we chose a Reed-Solomon code of dimension $32$ over $\mathbb{F}_{256}$ and, for the internal code, we chose the Reed-Muller code $[128, 8, 64]$ that we are going to duplicate between $2$ and $6$ times (i.e duplicating each bit to obtain codes of parameters $[256, 8, 128], [512, 8, 256], [786, 8, 384]$).

\medskip

\textbf{Decoding:} We perform maximum likelihood decoding on the internal code. 
This yields a vector of $\mathbb{F}_q^{n_e}$ that we then decode using an algebraic decoder for the Reed-Solomon code.

\medskip

\textbf{Decoding the internal Reed-Muller code:} The Reed-Muller code of order 1
can be decoded using a fast Hadamard transform (see chapter 14 of \cite{MWS77}).
The algorithm needs to be slightly adapted when decoding duplicated codes. For
example, if the Reed-Muller of length $128=2^7$ is duplicated three times, we create
the function $F: \mathbb{F}_2^{7} \rightarrow \{3, 1, -1, -3\}$ (which can be
thought of as a 128-tuple of symbols from $\{3, 1, -1, -3\}$)
by transforming every block of three bits $y_1y_2y_3$ of the received vector of
length $384$ to

$$ (-1)^{y_1} + (-1)^{y_2} + (-1)^{y_3}.$$

We then apply the Hadamard transform to the output of the function $F$. We take the maximum
value in $\hat{F}$ and $x \in \mathbb{F}_2^{128}$ that maximizes the value of $|\hat{F}|$. If $\hat{F}(x)$ is positive, then the closest codeword is $xG$ where $G$ is the generator matrix of the Hadamard code (without the all-one-vector). If $\hat{F}(x)$ is negative, then we need to add the all-one-vector to it.

\subsection{Decryption failure rate analysis}

We now consider the decoding failure rate of the concatenated code which also
corresponds to the decryption failure rate of the encryption scheme. We first
provide two bounds on the maximum likelihood decoding error probability of the
duplicated Reed-Muller code: a first simple union bound and a second more
accurate one.
These bounds can then be plugged into the decoding error probability
for the bounded distance decoder of the Reed-Solomon code. 

\begin{proposition}{\bf [Simple Upper Bound for the DFR of the internal code]} \label{prop:interal_DFR}

Let $p$ be the transition probability of the binary symmetric channel. Then the DFR of a duplicated Reed-Muller code of dimension $8$ and minimal distance $d_i$ can be upper bounded by:

$$ p_i = 255 \sum\limits_{j=d_i/2}^{d_i} {d_i \choose j} p^j(1-p)^{d_i-j} $$
\end{proposition}

\proof

For any linear code $C$ of length $n$, when transmitting a codeword $\cv$, the
probability that the channel makes the received word $\yv$ at least as close to
a word $\cv' = \cv + \xv$ as $\cv$ (for $\xv$ a non-zero word of $C$ and $|\xv|$ the weight of $\xv$) is:

$$ \sum\limits_{j \geqslant |\xv|/2} {|\xv| \choose j} p^j(1-p)^{n-j}.$$

By the union bound applied on the different non-zero codewords $\xv$ of $C$, we obtain that the probability of a decryption failure can thus be upper bounded by:

$$ \sum\limits_{\xv \in C, \xv \neq 0} \sum\limits_{j \geqslant |\xv|/2} {|\xv| \choose j} p^j(1-p)^{n-j}$$

There are 255  non-zero words in a [128,8,64] Reed-Muller code, 254 of weight 64 and one of weight 128. The contribution of the weight 128 vector is smaller than the weight 64 vectors, hence by applying the previous bound to duplicated Reed-Muller codes we obtain the result.
\qed

\medskip

\noindent \textbf{Better upper bound on the decoding error probability for the internal code.}
The previous simple bound pessimistically assumes that decoding fails when
more than one codeword minimizes the distance to the received
vector. The following bound improves the previous one by taking into account the
fact that decoding can still succeed with probability $1/2$ when exactly two
codewords minimize the distance to the received vector.

\begin{proposition}{ \bf [Improved Upper Bound for the DFR of the internal code]} \label{prop:better_interal_DFR}

Let $p$ be the transition probability of the binary symmetric channel. Then the DFR of a Reed-Muller code of dimension $8$ and minimal distance $d_i$ can be upper bounded by:

\begin{align}
p_i & = \frac{1}{2} 255 {d_i \choose d_i/2} p^{d_i/2}(1-p)^{d_i/2} \nonumber \\
& + 255 \sum\limits_{j=d_i/2+1}^{d_i} {d_i \choose j}p^j (1-p)^{d-j} \nonumber \\
& + \frac{1}{2} {255 \choose 2} \sum\limits_{j=0}^{d_i/2} {d_i/2 \choose j}^3
p^{d_i-j} (1-p)^{d_i/2+j} \nonumber
\end{align}
\end{proposition}

\proof

Let E be the decoding error event. Let $\ev$ be the error vector.

\begin{itemize}
\item Let $A$ be the event where the closest non-zero codeword $\cv$ to the
error is such that $d(\ev, \cv) = d(\ev, \mathbf{0}) = |\ev|$.
\item Let $B$ be the event where the closest non-zero codeword $\cv$ to the
error vector is such that $d(\ev, \cv) < |\ev|$.
\item Let $A' \subset A$ be the event where the closest non-zero codeword $\cv$
to the error vector is such that $d(\ev, \cv) = |\ev|$ and such a vector is
unique, meaning that for every $\cv' \in C, \cv' \neq \cv, \cv' \neq
\mathbf{0}$, we have $d(\ev, \cv') > |\ev|$.
\item Finally, let $A''$ be the event that is the complement of $A'$ in $A$,
meaning the event where the closest nonzero codeword $\cv$ to the error is at
distance $|\ev|$ from $\ev$, and there exists at least one codeword $\cv', \cv'
\neq \cv, \cv' \neq \mathbf{0}$, such that $d(\ev, \cv') = d(\ev, \cv) = |\ev|$.
\end{itemize}

The probability space is partitioned as $\Omega = A \cup B \cup C = A' \cup A'' \cup B \cup C$, where $C$ is the complement of $A \cup B$. When $C$ occurs, the decoder always decodes correctly, i.e. $P(E|C) = 0$. We therefore write:

$$P(E) = P(E|A')P(A') + P(E|A'')P(A'') + P(E|B)P(B)$$

When the event $A'$ occurs, the decoder chooses at random between the two closest codewords and is correct with probability $1/2$, i.e. $P(E|A') = 1/2$. We have $P(E|B) = 1$ and writing $P(E|A'') \leqslant 1$, we have:

\begin{align} \label{eq:PE}
P(E) & \leqslant \frac{1}{2}P(A') + P(A'') + P(B) \nonumber \\
     & = \frac{1}{2}(P(A') + P(A'')) + \frac{1}{2}P(A'') + P(B) \nonumber \\
P(E) & \leqslant \frac{1}{2}P(A) + \frac{1}{2}P(A'') + P(B)
\end{align}

Now we have the straightforward union bounds:

\begin{equation} \label{eq:PB}
P(B) \leqslant 255 \sum\limits_{j=d_i/2+1}^{d_i} {d_i \choose j}p^j (1-p)^{d-j}
\end{equation}

\begin{equation} \label{eq:PA}
P(A) \leqslant 255 {d_i \choose d_i/2} p^{d_i/2}(1-p)^{d_i/2}
\end{equation}

and it remains to find an upper bound on $P(A'')$.

We have:

$$P(A'') \leqslant \sum\limits_{\cv,\cv'} P(A_{\cv, \cv'})$$

where the sum is over pairs of distinct nonzero codewords and where:

$$A_{\cv, \cv'} = \{ d(\ev, \cv) = d(\ev, \cv') = |\ev| \}$$

This event is equivalent to the error meeting the supports of $\cv$ and $\cv'$ on exactly half their coordinates. All codewords except the all-one vector have weight $d_i$, and any two codewords of weight $d_i$ either have non-intersecting supports or intersect in exactly $d/2$ positions. $P(A_{\cv, \cv'})$ is largest when $\cv$ and $\cv'$ have weight $d$ and non-zero intersection. In this case we have:

$$P(A_{\cv, \cv'}) = \sum\limits_{j=0}^{d_i/2} {d_i/2 \choose j}^3 p^{d_i-j}
(1-p)^{d_i/2+j}$$

Hence

\begin{equation} \label{eq:PAsecond}
P(A'') \leqslant \sum\limits_{\cv,\cv'} P(A_{\cv, \cv'}) \leqslant {255 \choose
2} \sum\limits_{j=0}^{d_i/2} {d_i/2 \choose j}^3 p^{d_i-j} (1-p)^{d_i/2+j}
\end{equation}

Plugging \ref{eq:PA}, \ref{eq:PB} and \ref{eq:PAsecond} into \ref{eq:PE} we obtain the result. \qed

\begin{remark}
Propositions \ref{prop:interal_DFR} and \ref{prop:better_interal_DFR} give upper
bounds on the Decryption Failure Rate for the internal code. The smaller the
DFR, the closer the bounds become to the real value. We give a comparison of the
bounds from \ref{prop:interal_DFR} and \ref{prop:better_interal_DFR} and the
actual DFR for $[256, 8, 128]$, $[512, 8, 256]$ and $[768, 8, 384]$ duplicated
Reed-Muller using $p^\star$ values from actual parameters. Simulation results
are presented in Table~\ref{tab:RM}.
\end{remark}

{\small
\begin{table}[h]
\centering
\makebox[\textwidth][c]{
\begin{tabular}{|c|c|c|c|c|c|}
\hline
Security level & $p^\star$ & Reed-Muller code & DFR from \ref{prop:interal_DFR} & DFR from \ref{prop:better_interal_DFR} & Observed DFR\\
\hline
128 & 0.3196 & $[256, 8, 128]$ & -7.84 & -8.03 & -8.72\\
\hline
192 & 0.3535 & $[512, 8, 256]$ & -11.81 & -12.12 & -12.22\\
\hline
256 & 0.3728 & $[768, 8, 384]$ & -13.90 & -14.20 & -14.25\\
\hline
\end{tabular}
}
\caption{\label{tab:RM}Comparison between the observed Decryption Failure Rate
and the formula from proposition \ref{prop:interal_DFR}. Results are presented
as $\log_2(DFR)$.}
\end{table}
}

\begin{figure}[h!]
\centering
\includegraphics{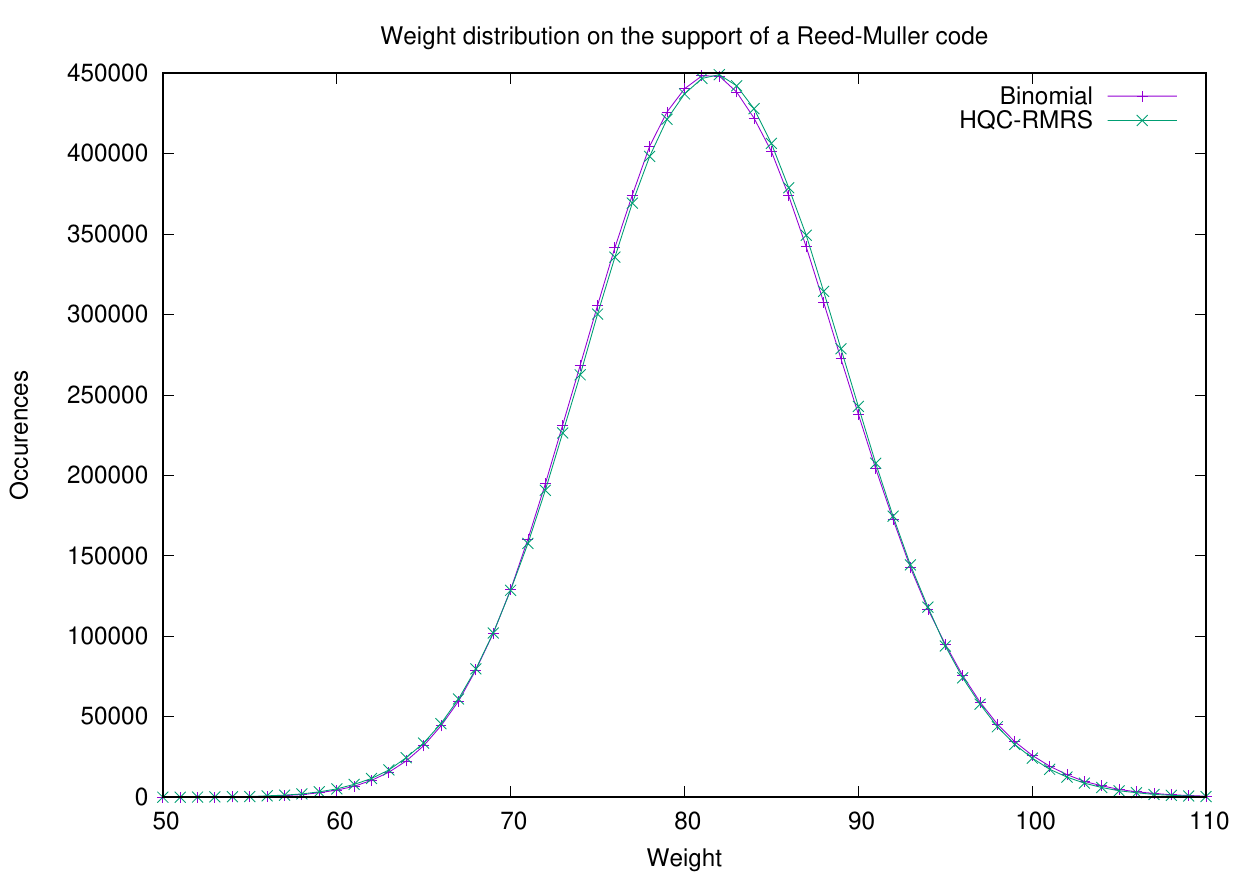}
\caption{The binomial distribution vs the actual weight distribution of the HQC
error vector restricted to the support of a Reed-Muller code. Parameters
correspond to parameter set II, and the support length is 256.}
\label{fig:binomial256}
\end{figure}

\begin{remark}
Propositions \ref{prop:interal_DFR} and \ref{prop:better_interal_DFR} have been
derived with a binary symmetric channel model for the distribution of the HQC error
vector restricted to the support of a (duplicated) Reed-Muller code. 
Figure~\ref{fig:binomial256} compares the actual weight distribution of
the error vector to the binomial distribution when restricted to this relatively
small number of bits. We observe that they are virtually identical, meaning that
a small proportion of HQC bits do behave as i.i.d Bernoulli variables.
\end{remark}

\begin{theorem}{\bf [Decryption Failure Rate of the concatenated code]}

Using a Reed-Solomon code $[n_e, k_e, d_e]_{\mathbb{F}_{256}}$ as the external code, the DFR of the concatenated code can be upper bounded by:

$$ \sum \limits_{l = \delta_e + 1}^{n_e} {n_e \choose l} p_i^l(1-p_i)^{n_e-l}$$

Where $d_e = 2\delta_e + 1$ and $p_i$ is defined as in Proposition\ref{prop:interal_DFR}.

\end{theorem}

\subsection{Simulation results}

\begin{figure}[h!]
\centering
\includegraphics{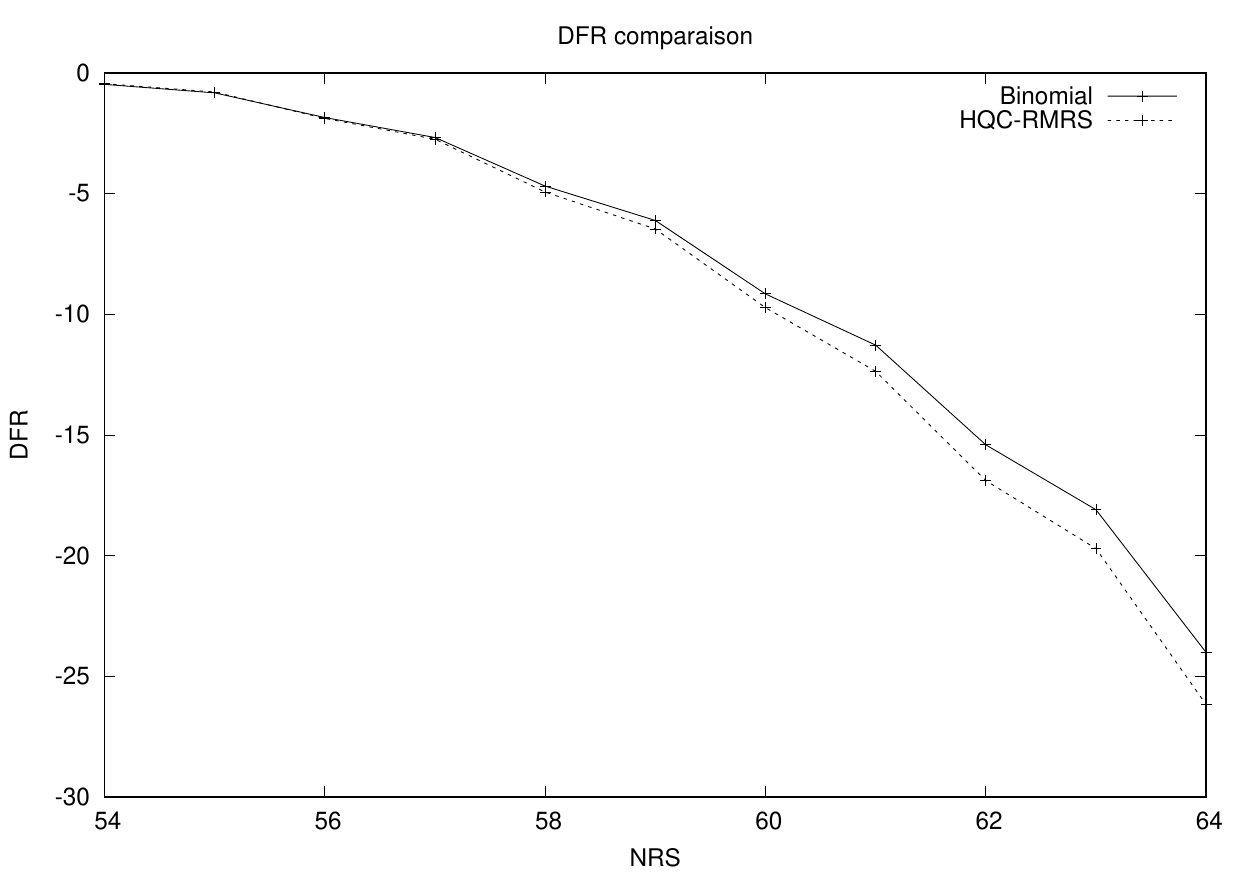}
\caption{Comparison of the Decryption Failure Rate of concatenated codes against
approximation by a binary symmetric channel and against HQC error vectors.
Parameters simulated are derived from those of HQC-RMRS for 128 security bits:
$w=67, w_r=w_e=77,$ a $[256,8,128]$ duplicated Reed-Muller code for internal
code and a $[\text{\tt NRS},32]$ Reed-Solomon code for external code.} 
\label{fig:binom_vs_hqc_concat}
\end{figure}

We tested the Decryption Failure rate of the concatenated codes against both
binary symmetric channels and HQC vectors. 
For Reed-Muller codes, rather than considering the upper bound approximation we
effectively decoded the code, which means than in practice the upper 
bound that we use for our theoretical DFR, is greater than what is obtained in
the simulations. Simulation results are presented on Figure~\ref{fig:binom_vs_hqc_concat}. These results show that the DFR of the encryption scheme is smaller than the simulated error with a binomial distribution which is itself smaller than the DFR derived from the bound on the internal duplicated Reed-Muller code.

\subsection{Proposed parameters}

From the DFR analysis we derive new parameters for the HQC-RMRS cryptosystem.
These are described on Figure \ref{fig:new_parameters}.

{\small
\begin{figure}[h!]
\centering
\makebox[\textwidth][c]{
\begin{tabular}{|c|c|c|c|c|c|c|c|c|}
\hline
Instance & security & $w$ & $w_{\mathbf{e}} = w_{\mathbf{r}}$ & Reed-Muller & Reed Solomon & $n$ & $DFR$ & Gain over \cite{ABDGZ18}\\
\hline
HQC-RMRS-128 & 128 & 67 & 77 & $[256, 8, 128]$ & $[80, 32, 49]$ & 20,533 & $< 2^{-128}$ & 16.8\%\\
\hline
HQC-RMRS-192 & 192 & 101 & 117 & $[512, 8, 256]$ & $[76, 32, 45]$ & 38,923 & $< 2^{-192}$ & 16.7\%\\
\hline
HQC-RMRS-256 & 256 & 133 & 153 & $[768, 8, 384]$ & $[78, 32, 47]$ & 59,957 & $< 2^{-256}$ & 15.4\%\\
\hline
\end{tabular}
}
\caption{New proposed parameters for the HQC-RMRS cryptosystem (security is in bits).}
\label{fig:new_parameters}
\end{figure}
}


\section{Conclusion}

In Section \ref{sec:error_distribution} we presented a better analysis of the
error weight distribution for HQC, which leads to a better DFR estimation. This
can be used to reduce the size of the parameters, no matter what family of codes
is used for decoding. In Section \ref{sec:concatenated} we propose using a
concatenation of Reed-Muller and Reed-Solomon codes and we provide an upper
bound on the DFR in this setting. This family allows us to reduce the public key
and ciphertext sizes by about $17\%$ when compared to the tensor product of BCH
and repetition codes (when considering the same error weight distribution).


\begin{thebibliography}{1}

\bibitem{ABDGZ18}
Carlos {Aguilar-Melchor}, Olivier Blazy, {Jean-Christophe} Deneuville, Philippe
  Gaborit, and Gilles Z{\'e}mor.
\newblock Efficient encryption from random quasi-cyclic codes.
\newblock {\em IEEE Transactions on Information Theory}, 64(5):3927--3943,
  2018.

\bibitem{MWS77}
FJ~MacWilliams and N.J.A. Sloane.
\newblock {\em The theory of error-correcting codes}.
\newblock North-Holland, 1977.

\end{thebibliography}
\end{document}